\documentclass[%
 reprint,
 superscriptaddress,
 nofootinbib,
 amsmath,amssymb,
 aps,
]{revtex4-2}

\usepackage{graphicx}
\usepackage{dcolumn}
\usepackage{bm}
\usepackage{hyperref}
\usepackage{xcolor}

\newcommand{\mnras}{MNRAS}
\newcommand{\aap}{A\&A}

\newcommand{\apjs}{ApJS}
\newcommand{\apjl}{ApJL}

\newcommand{\jcap}{JCAP}

\newcommand{\nphysa}{Nucl. Phys. A}

\newcommand{\msun}{M_\odot}
\newcommand{\hubble}{\ensuremath{H_0}}

\newcommand{\decel}{\ensuremath{q_0}}
\newcommand{\jerk}{\ensuremath{j_0}}

\newcommand{\vobs}{\ensuremath{\hat{v}}}

\newcommand{\zobs}{\ensuremath{\hat{z}}}
\newcommand{\zmax}{\ensuremath{z_{\rm max}}}
\newcommand{\prob}{\ensuremath{{\rm P}}}
\newcommand{\normal}{{\rm{N}}}

\newcommand{\nexp}{\bar{N}}
\newcommand{\abh}{a_{\rm BH}}
\newcommand{\ans}{a_{\rm NS}}
\newcommand{\mbh}{m_{\rm BH}}
\newcommand{\mns}{m_{\rm NS}}

\newcommand{\uniform}{{\rm U}}
\newcommand{\tobs}{t_{\rm obs}}
\newcommand{\fobs}{\Delta_{\rm obs}}
\newcommand{\rate}{\Gamma}
\newcommand{\step}{\Theta}

\newcommand{\snrmin}{\rho_*}
\newcommand{\mejmin}{m_{\rm ej}^*}
\newcommand{\dgw}{\hat{\bm{x}}}
\newcommand{\kmsmpc}{\ensuremath{{\rm km\,s^{-1}\,Mpc^{-1}}}}
\newcommand{\kms}{\ensuremath{{\rm km\,s^{-1}}}}

\newcommand{\yr}{\ensuremath{{\rm yr}}}
\newcommand{\yrgpc}{\ensuremath{{\rm yr^{-1}\,Gpc^{-3}}}}
\newcommand{\planck}{{\it Planck}}

\newcommand{\lcdm}{$\Lambda$CDM}

\newcommand{\seobnr}{\texttt{SEOBNR}}
\newcommand{\seobnrfull}{\texttt{SEOBNRv4\_ROM\_NRTidalv2\_NSBH}}
\newcommand{\imrp}{\texttt{IMRPhenom}}
\newcommand{\imrpfull}{\texttt{IMRPhenomPv2\_NRTidal}}

\begin{document}

\title{Prospects for Measuring the Hubble Constant with Neutron-Star-Black-Hole Mergers}

\author{Stephen M. Feeney}
\affiliation{Department of Physics \& Astronomy, University College London, Gower Street, London WC1E 6BT, UK}
\author{Hiranya V. Peiris}
\affiliation{Department of Physics \& Astronomy, University College London, Gower Street, London WC1E 6BT, UK}
\affiliation{Oskar Klein Centre for Cosmoparticle Physics, Department of Physics,
Stockholm University, AlbaNova, Stockholm SE-106 91, Sweden}
\author{Samaya M. Nissanke}
\affiliation{GRAPPA, Anton Pannekoek Institute for Astronomy and Institute of High-Energy Physics, University of Amsterdam, Science Park 904, 1098 XH Amsterdam, The Netherlands}
\affiliation{Nikhef, Science Park 105, 1098 XG Amsterdam, The Netherlands}
\author{Daniel J. Mortlock}
\affiliation{Astrophysics Group, Imperial College London, Blackett Laboratory, Prince Consort Road, London SW7 2AZ, UK}
\affiliation{Department of Mathematics, Imperial College London, London SW7 2AZ, UK}
\affiliation{Department of Astronomy, Stockholm University, AlbaNova, SE-10691 Stockholm, Sweden}

\date{\today}


\begin{abstract}

Gravitational wave (GW) and electromagnetic (EM) observations of neutron-star-black-hole (NSBH) mergers can provide precise local measurements of the Hubble constant (\hubble), ideal for resolving the current \hubble\ tension. We perform end-to-end analyses of realistic populations of simulated NSBHs, incorporating both GW and EM selection for the first time. We show that NSBHs could achieve unbiased 1.5--2.4\% precision \hubble\ estimates by 2030. The achievable precision is strongly affected by the details of spin precession and tidal disruption, highlighting the need for improved modeling of NSBH mergers.
\end{abstract}


\maketitle


\textbf{\emph{ Introduction.}} The current expansion rate of the Universe -- the Hubble constant, $H_0$ -- is at the heart of a significant cosmological controversy. Direct measurements in the local Universe by the SH0ES team's Cepheid-supernova distance ladder~\cite{Riess_etal:2019} find $\hubble = 74.03 \pm 1.42 \,\kmsmpc$. This is discrepant at the 4.4-$\sigma$ level from the $67.36 \pm 0.54\,\kmsmpc$ value inferred from the \planck\ satellite's observations of the cosmic microwave background (CMB) anisotropies, assuming the standard flat cosmological model~\cite{Planck_VI:2018}.

There are two potential explanations for this discrepancy, the most exciting of which derives from the model-dependence of the CMB constraint: could the discrepancy be due to physics beyond the standard model? Despite extensive effort (e.g., Refs.~\cite{Knox_Millea:2020,Vagnozzi:2020}), consensus on a compelling theoretical explanation has not been reached. The more prosaic explanation posits undiagnosed systematic errors or underestimated uncertainties; however, despite multiple investigations of both the distance ladder~\cite{Efstathiou:2014,*Rigault_etal:2015,*Jones_etal:2015,*Cardona_etal:2016,*Zhang_etal:2017,*Follin_Knox:2017,*Feeney_etal:2017,*Wu_Huterer:2017,*Dhawan_etal:2017,*Bengaly_etal:2018,*Rigault_etal:2018,*Jones_etal:2018,*Riess_etal:2020,*Efstathiou:2020} and CMB~\cite{Spergel_etal:2015,*Addison_etal:2016,*Obied_etal:2017,*Calabrese_etal:2017,*Efstathiou_Gratton:2019,*Motloch_Hu:2020,*ACT:2020} datasets, no study has found incontrovertible evidence warranting a change of conclusions.

In the absence of conclusive evidence of systematic errors or consensus on an extended model, independent verifications of the two central measurements offer a promising route to resolving the tension. Independent verification of the CMB anisotropy constraints comes from recent inverse distance ladder datasets~\cite{Addison_etal:2017,*DES_H_0:2017,*Philcox_etal:2020}. Local verification has, however, proven more challenging, with some alternative analyses supporting the SH0ES team's findings~\cite{Yuan_etal:2019,*Huang_etal:2020,*H0LICOW_XIII:2020,*TDCOSMO_I:2020,*Pesce_etal:2020} and others providing contradictory conclusions of varying significance~\cite{Freedman_etal:2019,*Freedman_etal:2020,*TDCOSMO_IV:2020,*Boruah_etal:2020}, in some cases using the same data.

A direct, completely independent local measurement with percent-level precision is therefore needed to resolve the \hubble\ tension. Combined gravitational-wave (GW) and electromagnetic (EM) observations of nearby compact-object mergers are ideal candidates to provide that measurement, yielding \hubble\ estimates that depend on general relativity alone~\cite{Schutz:1986,Holz_Hughes:2005,Dalal:2006,Nissanke_etal:2010,Taylor_etal:2012,Messenger_Read:2012,Nissanke_etal:2013,Oguri:2016,delPozzo:2017,Abbott_etal:2017a,Seto:2018,Chen_etal:2018,Fishbach_etal:2018,Feeney_etal:2018,Mortlock_etal:2019,Gray_etal:2019,Soares-Santos_etal:2019,Palmese_etal:2020,Mukherjee_Wandelt:2018,Mukherjee_etal:2020a,Vasylyev_Filippenko:2020,Chen_etal:2020,Gayathri_etal:2020,Mukherjee_etal:2020b}. Thanks to their accompanying EM emission, the utility of binary neutron star (BNS) mergers is well established~\cite{Dalal:2006,Nissanke_etal:2010,Taylor_etal:2012,Messenger_Read:2012,Nissanke_etal:2013,Oguri:2016,delPozzo:2017,Abbott_etal:2017a,Seto:2018,Chen_etal:2018,Fishbach_etal:2018,Feeney_etal:2018,Mortlock_etal:2019,Gray_etal:2019}, but less attention has been paid to the potential contribution of as-yet undiscovered neutron-star-black-hole (NSBH) mergers with EM counterparts~\cite{Nissanke_etal:2010,Nissanke_etal:2013,Vitale_Chen:2018}. Using idealized, fixed-signal-to-noise simulations at indicative parameter values,~\citet{Vitale_Chen:2018} recently showed that catalogs of GW-selected NSBH observations may constrain \hubble\ as well as BNSs, depending on the relative merger rates and BH spins. In particular, they showed explicitly that luminosity distance estimates could improve, as misaligned BH spins induce spin precession, helping break the degeneracy between the luminosity distance and inclination angle for some NSBH systems.

Here, we determine the \hubble\ constraints realistic NSBH samples will achieve, by performing end-to-end analyses of simulated NSBH samples incorporating fully specified parent populations, combined GW and EM selection, and a complete noise treatment. We use state of the art GW waveforms~\cite{Dietrich_etal:2019, Matas_etal:2020} and EM outflow models~\cite{Foucart_etal:2018}, both calibrated to a suite of numerical relativity simulations, for our GW and EM signals, and focus on the ``A+'' era of the mid-to-late 2020s, assuming an expanded GW network including LIGO India and KAGRA.


\begin{figure*}[ht!]
\includegraphics[width=18cm]{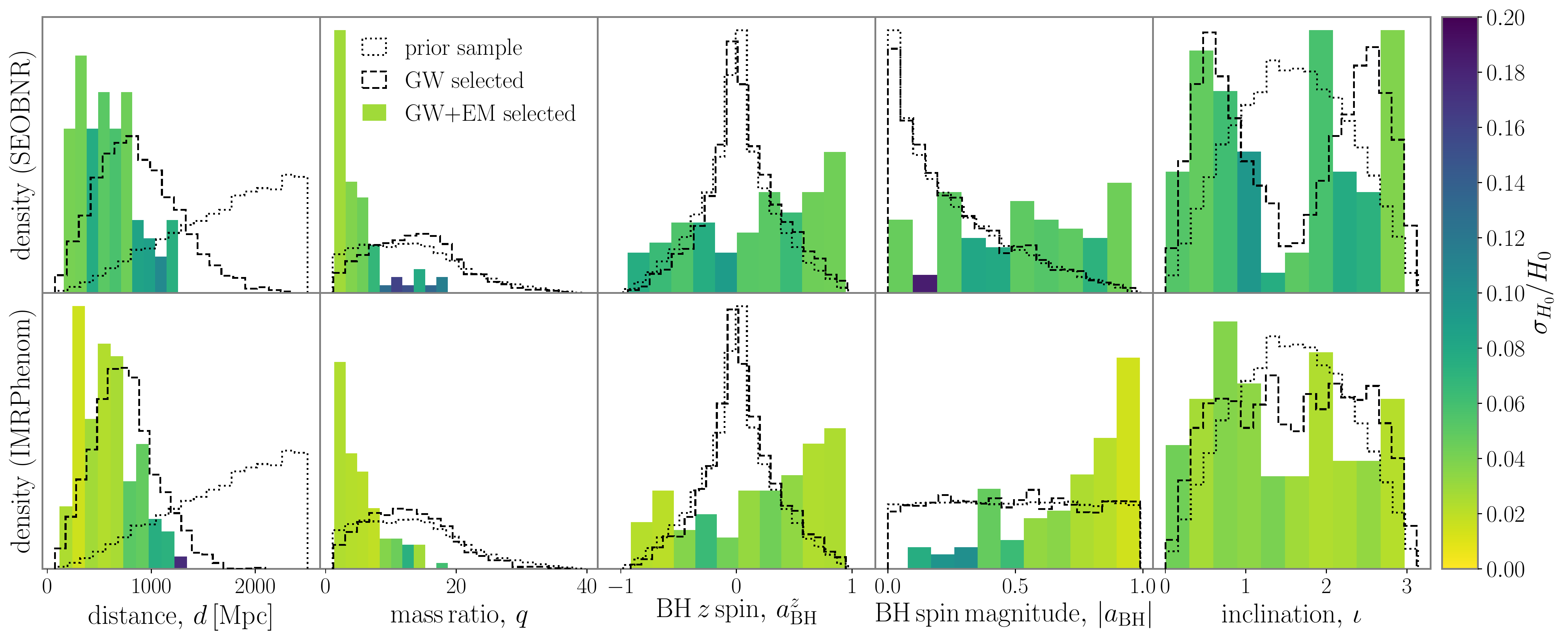}
\caption{Distributions of a subset of parameters from our \seobnr\ (top) and \imrp\ (bottom) samples, as drawn from the prior (dotted), selected by GW SNR (dashed) and selected by GW and EM emission (colored histograms). The bins are colored by the fractional \hubble\ uncertainty the mergers within the bin achieve: the yellowest/lightest bins are most informative. \label{fig:pops}}
\end{figure*}

\textbf{\emph{Simulations.}} In this work we simulate the results of a circa-2025 GW detector network, consisting of LIGO A+, Virgo AdV+, KAGRA and LIGO India~\cite{Abbott_etal:2013,LVCnoise} observing for $\tobs = 5\,\yr$ with duty cycle of $\fobs=0.5$. We assume a constant rest-frame  NSBH merger rate $\Gamma_{\rm fid} = 610$ \yrgpc\ (corresponding to the 90\% upper limit of Ref.~\cite{Ligo:2018}), and cosmological parameters matching Ref.~\cite{Planck_VI:2018}, with $\hubble=67.36\,\kmsmpc$ and $\decel=-0.527$. Each simulation proceeds by drawing the total number of mergers from a Poisson distribution with mean $\lambda = \fobs \, \tobs \, V \, \Gamma_{\rm fid} $, where $V$ is the redshifted volume (Eq.~\ref{eq:volume}) calculated using a third-order cosmographic co-moving volume element (Eq.~\ref{eq:volel}). To reduce computation time, the volume integral can be truncated at some redshift $\zmax$ where there is negligible probability of even the loudest merger being detected: we find that $\zmax=0.44$ suffices for our setting. For our fiducial parameter set, the mean number of mergers $\lambda = 25160$; our particular Poisson draw yields a pre-selection total of 25241.

For each merger, we draw a cosmological redshift, $z$, assuming a constant source-frame rate (see, e.g., Eq. 28 of Ref.~\cite{Mortlock_etal:2019}), along with an isotropically-distributed angular sky position, inclination angle, phase and polarization angle. We draw uniformly distributed BH masses from $\prob(\mbh) = \uniform(2.5\msun, 40\msun)$, taking the upper limit from low-metallicity\footnote{Using solar metallicities would reduce this upper limit to $12\,\msun$.} binary population synthesis simulations~\cite{Kruckow_etal:2018} and extending to low masses to reflect the detection of objects in the purported NS/BH mass gap~\cite{LVC:2020O3acat}. NS masses are drawn from $\prob(\mns) = \uniform(1\msun, 2.42\msun)$, with an upper limit chosen to match that of the DD2 equation of state (EOS)~\cite{Typel_etal:2010,*Hempel_etal:2010,*Fischer_etal:2014}. Dimensionless BH and NS spin magnitudes are drawn from the uniform distributions $\prob(\abh) = \uniform(0, 0.99)$ and $\prob(\ans) = \uniform(0, 0.05)$ and are assumed to be isotropically oriented. Following Refs.~\cite{Foucart:2012,Foucart_etal:2018}, we use the component masses, NS compactness and BH spins to calculate the baryonic mass ejected by each merger. This formula requires the assumption of a NS EOS (we again use DD2) and has been calibrated using simulations without precession due to misaligned BH spins. We use the same EOS to calculate tidal deformabilities for the NSs, and set the BH deformabilities to zero~\cite{Binnington_Poisson:2009,*Pani_etal:2015,*Cardoso_etal:2017,*Chia:2020}. Finally, we generate a peculiar velocity $v$ for each merger from a zero-mean normal with a standard deviation of $500\,\kms$.

With the NSBH parameters in hand, we generate mock data for each merger and apply our selection criteria.  To determine the impact of different physical effects on our results, we simulate two populations using different waveform approximants: the BNS-calibrated \imrpfull~\cite{Dietrich_etal:2019} and NSBH-specific \seobnrfull~\cite{Matas_etal:2020} (hereafter \imrp\ and \seobnr). We refer the reader to the Supplemental Material for a complete discussion of the differences between the two waveforms. The \seobnr\ waveform requires aligned or anti-aligned spins, so we set the transverse NS and BH spins to zero after sampling them isotropically (mimicking, in some sense, spins becoming aligned over time). For each merger, we generate a 32-second segment of noisy\footnote{Using spectra from Ref.~\cite{LVCnoise}.} GW data $\dgw$ per waveform using the same random seed and a frequency range of 20--2048 Hz, considering it detected if the network signal-to-noise ratio (SNR) is at least $\snrmin = 12$. We assume that the GW detectors operate in concert with an EM followup program capable of detecting all mergers with ejecta mass greater than $\mejmin = 0.01\,M_\odot$, modeling for the first time a hybrid GW-EM selection function for NSBHs. This ejecta material is assumed to produce EM emission in the form of a gamma-ray burst, kilonova and/or afterglow, as opposed to the ``battery'' effect of Ref.~\cite{Mingarelli_etal:2015}. Finally, we generate noisy measured redshifts and peculiar velocities by drawing from $\prob(\zobs|z)=\normal(z,0.001)$ and $\prob(\vobs|v)=\normal(v,200\,\kms)$, respectively. Of the 25241 simulated mergers, 2477 (2954) are detected in GWs using the \imrp\ (\seobnr) waveform, 99 (75) of which have sufficient ejecta to be detected in EM; 62 appear in both samples. The SNRs for \seobnr\ waveforms are, on average, 5.9\% larger than their \imrp\ counterparts, resulting in the GW-detected \seobnr\ sample containing $\sim$500 more objects.\footnote{We hypothesise that this is due to the effects of generic spin precession on the \imrp\ population. Differences in the lengths of GW signals within the detector frequency bands due to the two waveforms' different merger frequencies would tend to boost the \imrp\ SNRs.}  Setting the transverse spins to zero for use with the \seobnr\ waveform, however, has the side-effect of reducing the typical ejecta mass~\cite{Foucart_etal:2018} and hence the final GW+EM-detected sample.

The impact of our selection function is illustrated in Fig.~\ref{fig:pops}, in which we plot histograms of our full population (dotted lines), GW-selected events (dashed lines) and GW+EM-selected mergers (colored bars) for a subset of our parameters. The prior curves are identical in both cases apart from the BH spin magnitudes, where zeroing the transverse spins has made the \seobnr\ population's distribution non-uniform. The primary impact of the GW SNR threshold is, as expected, to select nearby mergers; it also imparts very slight preferences for low mass ratios and prograde BH $z$ spins~\cite{Ng_etal:2018}. It is interesting to note that the GW-selected \seobnr\ distance distribution is broader than that of \imrp\ and peaked at slightly higher distances: this is a direct consequence of the \seobnr\ injections' systematically higher SNRs. Further, the presence of spin precession permits the detection of more edge-on \imrp\ waveforms.

The ejecta-mass threshold (i.e., EM selection) strongly impacts the observed distributions. The GW+EM-detected distributions are shifted to even smaller distances, particularly for the \seobnr\ waveform, as the low-mass-ratio systems which produce significant ejecta mass can only be detected nearby. There is a very strong preference for mass ratios under 10 (again, especially so for \seobnr)\footnote{Populations with $\sim$solar metallicities produce NSBHs with lower BH masses~\cite{Kruckow_etal:2018} and hence more GW+EM-detectable mergers.} and large spins~\cite{Foucart:2012,Foucart_etal:2018}, and the preference for positive $z$ spins is much more pronounced. As expected from Refs~\cite{Foucart:2012,Mingarelli_etal:2015,Foucart_etal:2018}, the bulk of detected systems have BH masses below 10 $\msun$. The differences between the two waveforms' GW+EM-selected distributions are slightly obfuscated by the small sample sizes, but the \seobnr\ sample is shifted towards lower distances and mass ratios. As the \seobnr\ mergers' BH spin magnitudes are smaller than their \imrp\ counterparts', they require smaller mass ratios to produce significant ejecta~\cite{Foucart_etal:2018}, and the resulting systems are harder to detect at distance. Our implementation of EM selection captures the current best understanding of the dependence on ejecta mass, but we note that a fully self-consistent model of EM selection does not yet exist. This selection does not, for example, incorporate any viewing-angle dependence (see Ref.~\cite{Chen:2020} for a treatment in BNS mergers) or EM survey selection effects~\cite[e.g.,][]{Rosswog_etal:2017,Scolnic_etal:2018,Cowperthwaite_etal:2019,Setzer_etal:2019}.


\textbf{\emph{Methods.}} The probabilistic inference of the Hubble constant from catalogs of compact object mergers has been described in detail in the literature~\cite{Schutz:1986,Dalal:2006,Nissanke_etal:2010,Taylor_etal:2012,Nissanke_etal:2013,Abbott_etal:2017a,Chen_etal:2018,Fishbach_etal:2018,Feeney_etal:2018,Mandel_etal:2018,Gray_etal:2019,Mortlock_etal:2019,Vitale_etal:2020}. In the following, we adopt a slight variant of the formalism set out in Ref.~\cite{Mortlock_etal:2019}, whose Fig. 9 depicts a network diagram for the model we use to describe the data\footnote{The only addition required to the network diagram of Ref.~\cite{Mortlock_etal:2019} is the dependence of the selection, $S$, on an intrinsic parameter: the merger's ejecta mass.}. The posterior we evaluate is defined in Eq.~\ref{eq:posterior}.

We infer the parameters of this model in two parts, using two sampling methods. First, we process each merger individually in order to obtain the GW likelihoods marginalized over all parameters $\boldsymbol{\theta}_i$ other than the luminosity distance $d$ to the merger ($\boldsymbol{\theta}_i$ here comprising the $i^{\rm th}$ merger's component masses, spin magnitudes and orientations where used, inclination, polarization angle, NS tidal deformability, and time and phase at coalescence). We adopt priors identical to the distributions used in the generative model for all parameters other than the masses. Convergence is greatly improved by sampling chirp masses and mass ratios instead of component masses, and we therefore sample using interim priors that are uniform in these parameters (over the ranges permitted by our component-mass extrema), before importance-sampling the outputs to reinstate our desired component-mass priors. The marginal GW likelihoods are sampled with the \texttt{pypolychord} nested sampler~\cite{Handley_etal:2015a,*Handley_etal:2015b}, wrapped by \texttt{bilby}~\cite{Ashton_etal:2019}, using 1000 live points and \texttt{bilby}'s \texttt{marginalize\_phase}, \texttt{\_time} and \texttt{\_distance} settings. Each 15 (11)-dimensional \imrp\ (\seobnr) sampling run takes 6-14 (4-6) days to complete on one Intel Xeon 2.7 GHz CPU.

\begin{figure*}[ht!]
\includegraphics[width=18cm]{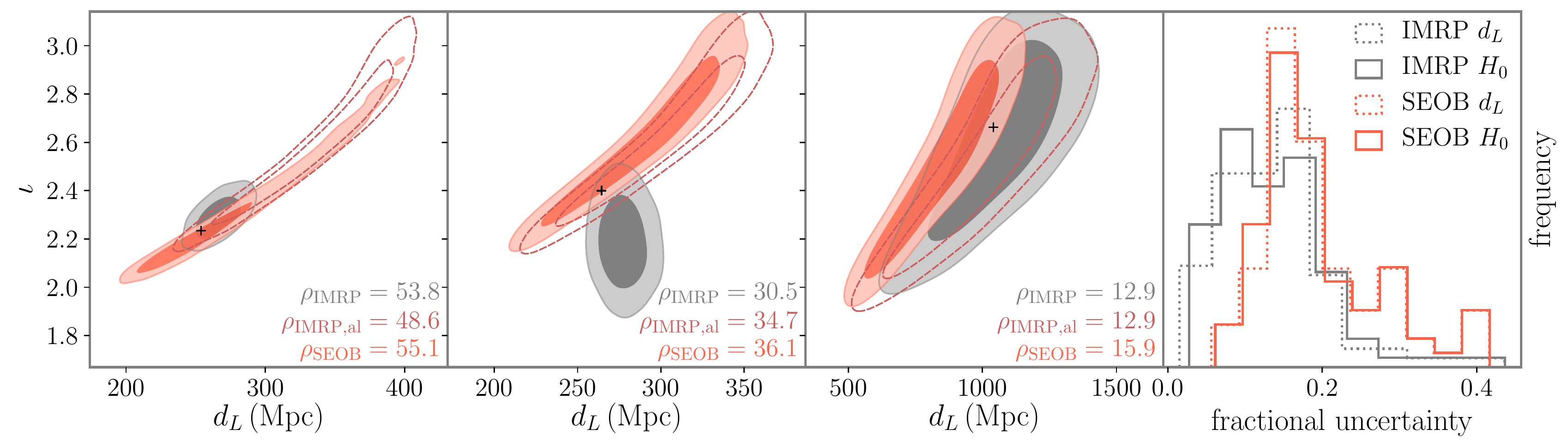}
\caption{Left three panels: distance and inclination posteriors for a selection of mergers, simulated and sampled using the \imrp\ waveform with precessing (grey filled) and aligned (dark red dashed) spins, and using \seobnr\ with aligned spins (red). The selection includes the highest-SNR merger common to both catalogs (left) and the \imrp\ merger whose BH spin is closest to being aligned (second from right). Right: distributions of fractional uncertainties on luminosity distance (dotted) and \hubble\ (solid) from individual mergers from our \imrp\ (grey) and \seobnr\ (red) NSBH catalogs. \label{fig:waveforms}}
\end{figure*}

Given the marginal GW likelihoods, we use No-U-Turn Sampling as implemented by the \texttt{pystan} package~\cite{pystan} to infer the cosmological and population parameters. To connect to the cosmological parameters, we adopt a third-order distance-redshift relation matching our volume element~\cite{Visser:2004}, with \hubble\ and \decel\ allowed to vary but the jerk set to one. We assume a broad Gaussian prior on \hubble, $\prob(\hubble)=\normal(70,20)$ \kmsmpc, a truncated Gaussian prior on \decel, $\prob(\decel)=\step(\decel+2)\step(1-\decel)\normal(-0.5,0.5)$, and a log-uniform prior on the rate, $\prob(\rate) \propto 1/\rate$. To use \texttt{pystan}, we must be able to sample all parameters from analytic distributions. We therefore perform a Gaussian Mixture Model fit to each merger's marginal distance likelihood using \texttt{pomegranate}~\cite{Schreiber:2017}. We fit each likelihood with an integer grid of 2--10 mixture components, repeating 10 times at each grid point and selecting the best fit using the Akaike Information Criterion~\cite{Akaike:1974}.

Finally, we must evaluate the expected number of detected mergers $\nexp$ at each sampled value of the cosmological and population parameters. We do so by re-simulating the catalogs 100 times at each point of a 5$\times$5 grid in $\{H_0,q_0\}$ assuming our fiducial rate, $\Gamma_{\rm fid}$, and interpolating the results using a 2D fourth-order interpolation. The dependence on the sampled rate is captured by multiplying the interpolation coefficients by $\Gamma/\Gamma_{\rm fid}$. The resulting 153 (193)-dimensional \texttt{pystan} inference runs take less than a minute to generate 20,000 well converged samples on a 3.1 GHz Intel Core i7 CPU. The set of true redshifts and peculiar velocities are uninteresting for the purposes of cosmology inference, and we marginalize over these parameters when quoting the results below.


\textbf{\emph{Results.}} Processing the simulated \seobnr\ and \imrp\ catalogs through our two-stage inference pipeline produces the cosmology and population parameter posteriors shown in the Supplemental Material. In both cases, the recovered \hubble, \decel\ and rate posteriors are completely consistent with the input values, indicating, as expected, that the selection effects are correctly accounted for~\cite{Mortlock_etal:2019}. The 68\% credible intervals on the near-Gaussian $\hubble$ marginal posteriors are $68.8 \pm 1.6\,\kmsmpc$ for the \seobnr\ sample and $66.5 \pm 1.0\,\kmsmpc$ for \imrp. As the \imrp\ sample contains 99 objects to the \seobnr\ sample's 75, we should therefore expect that the \imrp\ sample's $\hubble$ posterior be roughly 13\% narrower than that of the \seobnr\ sample. The remaining reduction, therefore, reflects the ability for precessing spins to break the distance-inclination degeneracy~\cite[e.g.,][]{Apostolatos_etal:1994,Cutler_Flanagan:1994,Vecchio:2004,Chen_etal:2018}. This additional constraining power is equivalent to an approximate doubling of the catalog size.

The \hubble\ uncertainties we find for both waveforms are comparable to the current Cepheid-SN distance ladder precision~\cite{Riess_etal:2019}. NSBH populations -- should they produce EM counterparts and occur at rates roughly matching our assumptions -- will therefore strongly inform the outcome of the current \hubble\ tension, particularly when combined with accompanying BNS populations\footnote{We note that comparable \hubble\ precision had been hoped for by 2023 from catalogs of BNS mergers~\cite{Chen_etal:2018}. That timescale, however, now appears optimistic due to the lack of EM counterpart detections following GW170817.}, likely of comparable size~\cite{Chen_etal:2018,Feeney_etal:2018,Vitale_Chen:2018,Mortlock_etal:2019}. The mergers are also informative about the deceleration parameter, $q_0$, shrinking its uncertainty from 0.5 to 0.32 or 0.27, depending on the waveform. This further implies that NSBH catalogs will be able to begin constraining parameters such as the matter density and dark energy EOS (in the context of \lcdm\ and extended models), complementary to BBH results from higher redshifts~\cite{Farr_etal:2019,Chen_etal:2020,Mukherjee_etal:2020b}. The merger rates are recovered with roughly 10\% precision~\cite[e.g.,][]{Ligo:2018,Abbott_etal:2018,LVC:2020O3acat,LVC_O2_pop}.

To obtain a picture of the parameter combinations that are most important for the \hubble\ constraints, we return to Fig.~\ref{fig:pops}. Here, the colors of the histogram bins indicate the fractional \hubble\ uncertainty the mergers within each bin attain, with the most-constraining bins colored yellow and the least-constraining blue. For both waveforms, the bulk of the \hubble\ constraining power comes from mergers out to roughly 700 Mpc, not just the very nearest ($\sim$200 Mpc), loudest events. For the \imrp\ mergers, all mass ratio bins less than $\sim$8$\msun$ contribute equally, despite the frequency dropping rapidly with mass ratio. For the \seobnr\ mergers, the constraints are instead driven by the lowest-mass ratio events~\cite[e.g., ][]{Apostolatos_etal:1994,Cutler_Flanagan:1994}. From the \imrp\ spin panels, it is clear that that highest-spin events constrain \hubble\ most strongly, with the full \hubble\ constraint coming almost entirely from the highest-spin (and most populated) bin. The \seobnr\ constraint, on the other hand, is sourced by events with a broader range of spins, though this is likely driven by the prior.

We further highlight the importance of precession in breaking the distance-inclination degeneracy in Fig.~\ref{fig:waveforms}. In the first three panels we plot distance and inclination constraints for a selection of mergers when using the \seobnr\ (red, filled) and \imrp\ (grey, filled) waveforms. For the first two mergers, detected with high SNR, the long degeneracies present in their \seobnr\ posteriors are almost completely broken when using the \imrp\ waveform, in which the spins precess. We illustrate this point further by re-running the \imrp\ case assuming aligned spins, having set the transverse spins to zero. These results are overlaid as dashed dark red contours. Both distance-inclination degeneracies blow up, increasing the distance uncertainties by a factor of over three, with commensurate consequences for the mergers' ability to constrain $H_0$. In the third panel, we show equivalent posteriors for the \imrp\ merger whose BH spin is closest to being aligned: the effect of switching waveforms here is markedly reduced. The impact on the population level is clear from the right-hand panel of Fig.~\ref{fig:waveforms}, in which we plot the distributions of individual mergers' fractional errors on distance (dashed) and \hubble\ (solid) when using the \seobnr\ (red) and \imrp\ (grey) waveforms. The \seobnr\ distributions are shifted to significantly higher errors than their \imrp\ counterparts, despite the \seobnr\ mergers typically having higher SNRs. The smallest percentage error for any individual merger is 2.8\% for the \imrp\ case and 6.1\% for \seobnr; the medians are 13.2\% and 17.3\%, respectively. The \hubble\ constraints imparted by both ``golden'' and normal events are therefore stronger when spins precess significantly. Finally, we note from the lower limits of the dashed curves that peculiar velocity {\it and redshift} uncertainties strongly suppress the constraining power of the nearest and loudest events.


\textbf{\emph{Conclusions.}} In this Letter, we present the results of the first end-to-end inference of \hubble\ from realistic simulated catalogs of NSBH mergers incorporating GW and EM selection effects. The precision we should expect from such catalogs is very promising for resolving the current \hubble\ tension, with five years of A+ era observations yielding \hubble\ uncertainties of 1.5--2.4\%. We find, however, that the detailed physics of the NSBH waveforms strongly impacts the achievable precision. Using the \seobnrfull\ waveform with non-precessing BH spins results in boosted SNRs, and an increase of $\sim$500 GW-detected NSBHs. However, including precessing spins using the \imrpfull\ waveform markedly increases the typical ejecta mass and hence the number of combined GW+EM detections. Critically, precessing spins also break the distance-inclination degeneracy in the resulting GW parameter posteriors, yielding a significant improvement ($\sim$40\% after accounting for differing catalog sizes) in the resulting \hubble\ constraint. Our results strongly highlight the need for improved modeling of NSBH signals in both gravitational waves (see e.g.,~\cite{YHuang_etal:2020}) and the electromagnetic spectrum.

\textbf{\emph{Software.}} The Python simulation and inference software developed for this analysis are publicly available at \url{https://github.com/sfeeney/nsbh}.


\textbf{\emph{Acknowledgments.}} We thank Sukanta Bose for providing details on LIGO India, Nikhil Sarin and Greg Ashton for help with \texttt{bilby}, Will Handley for help with \texttt{pypolychord}, and Tanja Hinderer, Andrew Williamson, Francois Foucart, and Bastien DuBoeuf for useful discussions. SMF is supported by the Royal Society. HVP's work was partially supported the research environment grant ``Gravitational Radiation and Electromagnetic Astrophysical Transients (GREAT)'' funded by the Swedish Research Council (VR) under Dnr 2016-06012 and the research project grant ``Gravity Meets Light'' funded by the Knut and Alice Wallenberg Foundation Dnr KAW 2019.0112. SMN is grateful for financial support from the Nederlandse Organisatie voor Wetenschappelijk Onderzoek (NWO) through the VIDI and Projectruimte grants. HVP and DJM acknowledge the hospitality of the Aspen Center for Physics, which is supported by National Science Foundation grant PHY-1607611. The participation of HVP and DJM at the Aspen Center for Physics was supported by the Simons Foundation. This work used computing facilities provided by the UCL Cosmoparticle Initiative; and we thank the HPC systems manager Edd Edmondson for his dedicated support. We are deeply grateful to the \imrp\ and \seobnr\ waveform modelers for making these waveforms public, without which this work would not have been possible.

\textbf{\emph{Author contributions.}} {\bf SMF}: conceptualization, methodology, software, investigation, validation, writing (original draft preparation). {\bf HVP}: conceptualization, methodology, validation, writing (review \& editing), funding acquisition. {\bf SMN}: conceptualization, methodology, validation, writing (review \& editing). {\bf DJM}: conceptualization, methodology, writing (review \& editing).


\appendix

\section{Supplemental Material}

\textbf{\emph{Waveform Approximants.}} The waveform approximants used in this work are based on the two main methods that model fully the inspiral-merger-ringdown phases of binary black holes. Both models extend the analytically--derived inspiral to incorporate the merger and ringdown by calibrating the full waveforms with a suite of numerical relativity simulations. With these baseline binary black hole models in hand, the two approaches treat the finite-size effects of neutron stars by adding tidal deformation, using BNS numerical relativity simulations, and spin-quadrupole corrections to the GW phase evolution in the waveforms. Specifically, \imrp\ uses a phenomenological approach when extending the post-Newtonian approximation of the inspiral using numerical relativity simulations, which encapsulates the dominant effects of generic precessing binary black holes ~\cite[e.g.,][]{Ajith_etal:2007,*Ajith_etal:2008,*Hannametal:2014}. Importantly, \imrp\ models the merger frequency of the system with a taper for BNS systems only. In contrast, \seobnr\ is based on the Effective-One-Body (EOB) formalism to model the two body problem in General Relativity~\cite[e.g.,][]{Buonanno_Damour:1999, *Buonanno_Damour:2000, *Bohe_etal:2017, *Barausse_Buonanno:2010}. \seobnr\ specifically applies to NSBH systems where the BH spin is aligned or anti-aligned to the orbital angular momentum plane. In addition to the finite-size-induced GW phase correction, \seobnr\ automatically describes the merger frequency for mass ratios of 1--5 using numerical relativity simulations of NSBHs. An updated, NSBH-specific waveform, \texttt{IMRPhenomNSBH}, was released in Ref.~\cite{Thompson_etal:2020}; however, we do not use it here as it is only valid for BH spin magnitudes up to 0.5.

\begin{figure*}[ht!]
\includegraphics[width=8cm]{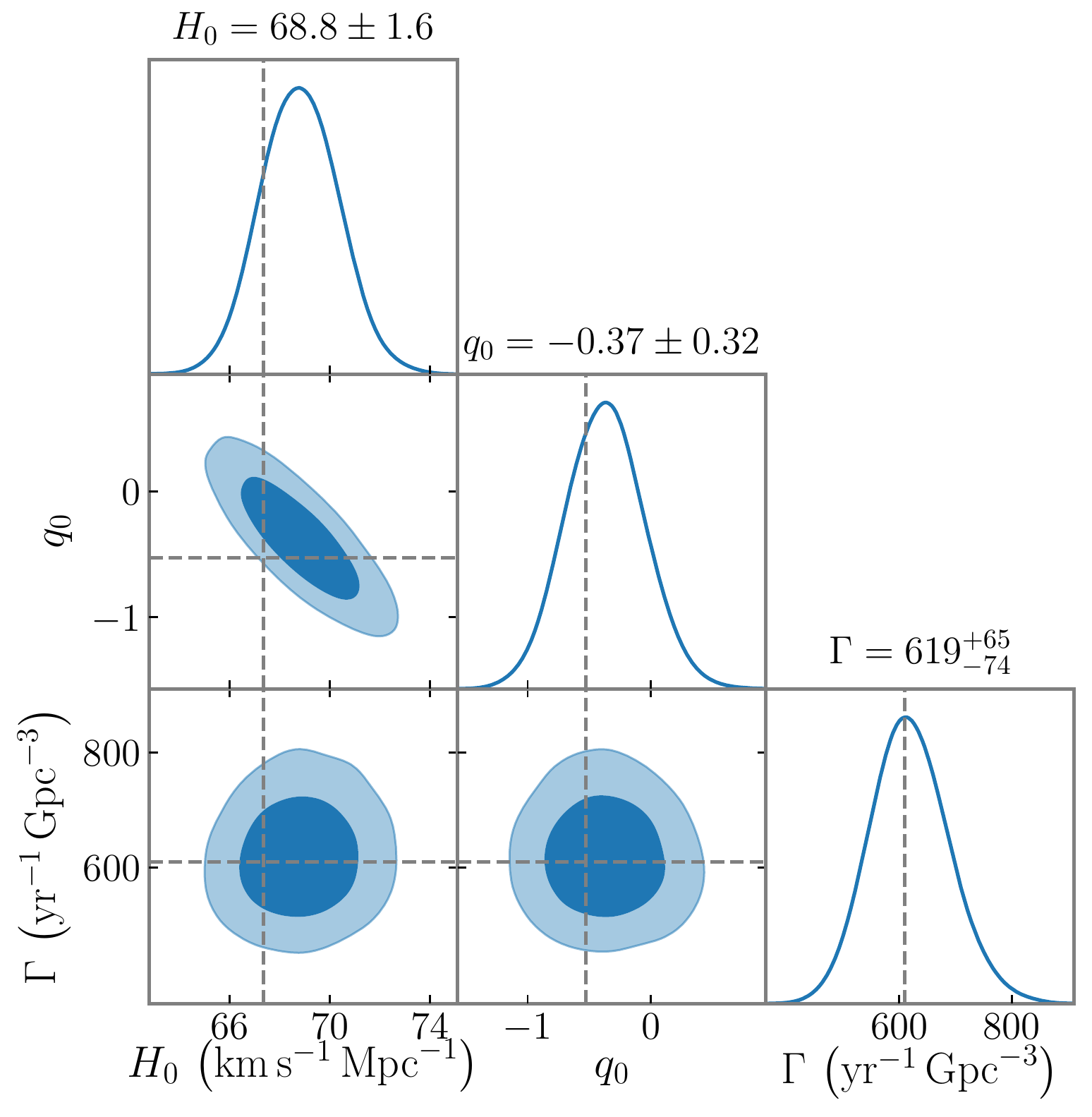}\includegraphics[width=8cm]{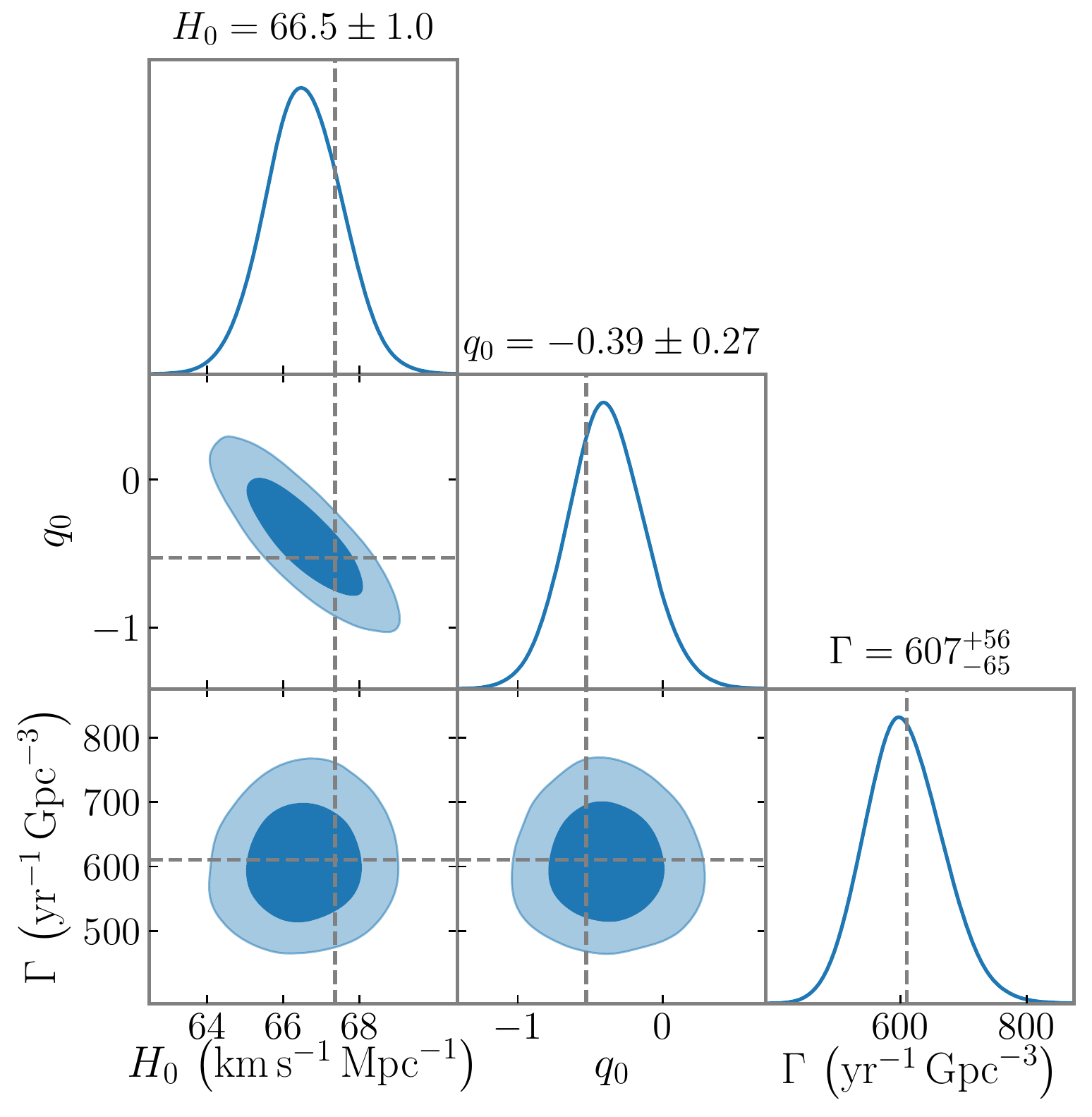}
\caption{Cosmological and population parameter posteriors inferred for the mock \seobnr\ (left) and \imrp\ (right) catalogs.\label{fig:cosmo}}
\end{figure*}

\textbf{\emph{Cosmography.}} We assume a cosmographic distance-redshift relation based on Taylor expanding the relevant quantities to third order in expansion parameters, i.e., the Hubble constant, \hubble, the deceleration parameter, \decel, and the jerk, $j_0$ (which we assume to be one throughout). The luminosity distance is then related to the expansion parameters by~\cite{Visser:2004}
\begin{equation}
d \simeq \frac{cz}{\hubble} \left[1 + \frac{1}{2} \left(1 - \decel \right) z + \frac{1}{6} \left(-1 + \decel - \jerk + 3 \decel^2 \right) z^2 \right],
\label{eq:distance}
\end{equation}
and the comoving volume element by
\begin{align}
\frac{{\rm d}V}{{\rm d}z} \simeq & 4\pi \frac{c^3 z^2}{\hubble^3} \times \label{eq:volel} \\
& \left[1 - 2 \left(1 + \decel \right) z + \frac{5}{12} \left(7 + 14 \decel - 2 \jerk + 9 \decel^2 \right) z^2 \right]. \nonumber
\end{align}
The redshifted volume is defined to be 
\begin{equation}
V = 4\pi \int_0^{\zmax} \frac{{\rm d}V}{{\rm d}z} \frac{{\rm d}z}{1+z}.
\label{eq:volume}
\end{equation}

\textbf{\emph{Posterior.}} Our two-stage Bayesian inference pipeline requires us to first evaluate, for each merger in turn, the GW likelihood marginalized over all parameters $\boldsymbol{\theta}_i$ other than the luminosity distance $d$ to the merger:
\begin{equation}
\prob(\dgw_i | d[ z_i, \hubble, \decel ]) = \int {\rm d}\boldsymbol{\theta}_i \prob(\boldsymbol{\theta}_i) \prob(\dgw_i | d[ z_i, \hubble, \decel ], \boldsymbol{\theta}_i).
\label{eq:marge_like}
\end{equation}
$\dgw_i$ here are the the $i^{\rm th}$ merger's GW strain observations, and $\boldsymbol{\theta}_i$ comprises the merger's component masses, spin magnitudes and orientations (if using precessing spins), inclination, polarization angle, NS tidal deformability, and time and phase at coalescence. With these in hand, the joint posterior of the cosmological and NSBH parameters is given by
\begin{align}
\prob & \left( \hubble, \decel, \Gamma, \{z, v\} | N, \left\{ \dgw, \hat{z}, \hat{v} \right\}, \snrmin, \mejmin \right) \propto \label{eq:posterior} \\
& \prob \left( \hubble, \decel, \Gamma \right) \exp \left( -\bar{N} \left[ \hubble, \decel, \Gamma, \snrmin, \mejmin \right] \right) \times \nonumber \\
& \prod_{i = 1}^{N} \frac{\Gamma}{1 + z_i} \frac{{\rm d}V}{{\rm d}z} \left[ \hubble, \decel \right] \prob (\dgw_i | d \left[ z_i, \hubble, \decel \right]) \prob(v_i) \prob(\hat{v}_i | v_i) \prob(\hat{z}_i | z_i), \nonumber
\end{align}
where $N$ and $\nexp$ are the actual and expected number of mergers detected, respectively, curly brackets denote sets of quantities and bold denotes per-merger vectors. The specific posteriors we obtain after processing our simulated catalogs are shown in Fig.~\ref{fig:cosmo}.


\bibliography{references}

\end{document}